\documentclass[twocolumn,showpacs,amsmath,amssymb,floatfix,prb,preprintnumbers,footinbib]{revtex4}
\usepackage{graphicx}
\usepackage{psfrag}
\usepackage{amsmath,amsfonts}
\usepackage{color}

\begin{document}
\title{Diagrammatic real-time approach to adiabatic pumping through metallic single-electron 
devices}

\author{Nina Winkler}
\affiliation{Theoretische Physik, Universit\"at Duisburg-Essen and CeNIDE, 47048 Duisburg, Germany}

\author{Michele Governale}
\affiliation{Theoretische Physik, Universit\"at Duisburg-Essen and CeNIDE, 47048 Duisburg, Germany}

\author{J\"urgen K\"onig}
\affiliation{Theoretische Physik, Universit\"at Duisburg-Essen and CeNIDE, 47048 Duisburg, Germany}

\date{\today}
\begin{abstract}
We present a real-time diagrammatic formalism to study adiabatic pumping in chains of tunnel-coupled metallic islands. This approach is based on an expansion to linear order in the frequency of the time-dependent parameters and on a systematic perturbation expansion in the tunnel-coupling strength.
We apply our formalism to single-island and double-island systems. In the single-island setup, we find that the first-order contribution in the tunnel-coupling strength is purely due to the renormalization of the charging-energy gap. In the double-island system, we investigate the transition between weak and strong pumping.
\end{abstract}



\maketitle

\section{Introduction}
The most common method to drive a current through a conductor is by applying a voltage. 
In nanostructures, due to the ease with which parameters can be tuned,  another possibility opens up: 
\textit{pumping.}
In an electron pump a  dc current can be generated in a nanoscale conductor, even at zero-bias voltage, by periodically changing some of its properties in time. If the time dependence of the parameters is slow compared to the internal time scales of the system, such as the dwell time of carriers, pumping is adiabatic. In this case, the pumped charge does not depend on the detailed time evolution of the pumping cycle but only on its area. 
Most of the theoretical efforts on pumping have been addressed to systems where the electron-electron interaction is weak and can, therefore, be neglected or treated in a mean-field approach.\cite{brouwer98,zhou99,buttiker01,makhlin01,buttiker02,entin02} For these weakly interacting nanoscale conductors, a well-established formalism for  pumping exists based on the dynamical-scattering approach to mesoscopic transport.\cite{buttiker94,brouwer98}
Recently, several works have investigated pumping through interacting systems.\cite{aleiner98,citro03,aono04,brouwer05,cota05,splett05,sela06,splett06,fioretto07,arrachea08,splett08} Most of these works have focused on few-electron quantum dots\cite{aono04,cota05,splett05,splett06,fioretto07,arrachea08,splett08} since these systems are the prototype for nanostructures with strong interaction effects. An open question regards how the pumping characteristics of an interacting quantum dot depends on the dot's single-particle spectrum. In order to address this problem, we have decided to study a quantum dot with a dense spectrum, i.e., with vanishing level spacing. Due to its finite density of states, such a quantum dot is referred to as a metallic island. 
Indeed, pumping in systems of metallic islands has been investigated experimentally already in the early 1990s.\cite{geerligs91,pothier92} Pothier {\it et al.} \cite{pothier92} realized a single-electron pump in a setup with two tunnel-coupled metallic islands in series. 
Pumping was obtained by changing in time the gate voltages of the two islands. The pumping cycle was chosen to enclose a degeneracy point, where three charge states of the double-island system are degenerate. 
This resulted in pumping of exactly one electron per cycle. To describe this experiment, it was assumed that the system relaxes always to the ground state and that this transition occurs via first-order tunneling. 
Although this theoretical explanation is perfectly suited to describe the pumping cycle of Ref.  \onlinecite{pothier92}, it would fail in describing weak pumping or pumping cycles nonencircling a degeneracy point. 

More experiments on charge pumping through nanostructures were performed on metallic multijunction systems,~\cite{martinis} double quantum dots in InAs quantum wires,~\cite{fuhrer}
or making use of surface acoustic waves.~\cite{fletcher,buitelaar, kaestner} 
Also spin pumping through a quantum dot was reported.~\cite{watson}

In the present paper,  we develop a diagrammatic real-time approach to
adiabatic pumping in systems consisting of metallic islands. To this aim, we follow the lines of Ref.~\onlinecite{splett06}, where adiabatic pumping through a single-level quantum dot was considered. 
Our approach relies on a systematic perturbative expansion in the tunnel-coupling strengths and therefore is restricted to weak tunnel coupling. 
In particular, it is not suitable to study the very-low-temperature regime, where high-order tunneling processes become relevant.
However, it treats the Coulomb interaction on the island exactly, allowing us to investigate effects due to strong interaction.

First, we apply our formalism to a single-island system consisting of one metallic island with Coulomb interaction,  tunnel coupled to two noninteracting leads. We compute the pumped charge up to first order in the tunnel-coupling strength for any pair of pumping parameters chosen among the charging-energy gap of the system and the left and the right tunnel-coupling strengths. As a result, we find that the contribution in first order is due solely to the renormalization of the charging-energy gap. For the case of pumping with the two tunnel-coupling strengths, this term becomes the dominant one. 
Furthermore, we consider pumping with the charging-energy gaps in a more complex system consisting of two tunnel-coupled metallic islands called the double-island system. We calculate the pumped charge through this system in lowest order in the tunnel-coupling strength and investigate the issue of pumped-charge quantization. With our technique we are able to describe the transition from weak to strong pumping, finding charge quantization for strong pumping. 

This paper is structured as follows. In Sec. II we present the model. 
The technique to compute the pumped current is introduced in Sec. III.
The results for single-island and double-island systems are presented in Secs.~IV and~V, respectively.
The conclusions are given in Sec.~VI. Details on the calculation of the diagrams can be found in Appendixes~A and B.

To keep all formulas transparent we set $\hbar =1$ throughout the paper.

\section{Model}
We consider a chain of $M$ tunnel-coupled  metallic islands. The leftmost and rightmost islands are tunnel coupled to metallic leads. The different metallic regions are labeled by the index $r$, where $r=0$ and $r=M+1$ correspond to the noninteracting metallic leads and $r=1,\dots, M$ to the metallic islands. The metallic islands, for typical experimental realizations, have a dense single-particle spectrum and can be treated as equilibrium reservoirs of fermionic degrees of freedom, described by a Fermi function. Each island is capacitively coupled to a gate voltage, which can be  used as a pumping parameter. 
The state of the chain is described by the number of excess charges $N_r$ on each island, $|\chi\rangle=|N_1,\dots,N_M\rangle$. 
The Hamiltonian of the system is  $H=H_{\mathrm{0}}+H_{\mathrm{tunn}}$. 
The operator $H_0$ describes the system in the absence of tunnel coupling
\begin{equation}
 H_{0} = \sum\limits_{r=0}^{M+1} \sum\limits_{k,n} \epsilon_{k nr} c^\dagger_{k nr} c_{k nr}+H_{\text{ch}}, 
\end{equation}
where $\epsilon_{{k nr}}$ are the single-particle energies, labeled by $k$ and by the transverse-channel index $n=1,\dots, N_{\rm chann}$. The spin degree of freedom is included in the sum over transverse channels. In the following, we will consider the many-channel case, $N_{\rm chann}\gg 1$.   The Coulomb interaction 
is taken into account by  $H_{\text{ch}}$, which describes the electrostatic energy of a given charge state $|\chi\rangle$.\cite{schoeller97} This charging energy depends on the gate voltages applied to each single island. 
The tunnel coupling is described by the tunneling Hamiltonian
\begin{eqnarray}
\label{htunn}
	H_{\text{tunn}} =& \sum\limits_{r=0}^{M} \sum\limits_{k,p,n}\left[ V_{r}(t) c^\dagger _{k nr} 	c_{p n \,r+1}\,e^{-i(\varphi_{r+1}-\varphi_r)} +\mbox{H.c.}\right] \, ,
\end{eqnarray}
with $\varphi_{0}=\varphi_{M+1}\equiv0$. For simplicity in Eq.~(\ref{htunn}) we have assumed that the tunnel-matrix elements $V_{r}$ for tunneling through barrier $r$ do not depend on the quantum numbers $k$ and $p$ or on the transverse-channel index $n$. Notice that the tunneling Hamiltonian Eq.~(\ref{htunn}) conserves the transverse-channel index. In Eq.~(\ref{htunn}) we have explicitly indicated that the tunnel-matrix elements can be time dependent.
The operator $\varphi_r$ is the canonical conjugate of the number operator $\hat{N}_r$ for the excess 
charges on island $r$, i.e., $[\varphi_r,\hat{N}_r]=i$. Therefore, $ e^{\pm i\varphi_{r}}$ are simply ladder operators for the charge on island $r$.
Finally, we characterize the junction $r$ between the two consecutive islands $r$ and $r+1$ by the tunnel-coupling strength
\begin{equation}
\begin{split}
\alpha_0^{r} (t)
&=\frac{R_\text{k}}{4 \pi^2 R_{r} (t)} \\
&=\sum_n N_r^n(0)\; N_{r+1}^n(0) \; V_{r} (t) V^*_r (t)\,, 
\end{split}
\end{equation}
where $R_{\text{k}}=h/e^2$ is the quantum resistance, $R_r$ the tunnel resistance of the junction,  $N_r^n(0)$ and $N_{r+1}^n(0)$ the densities of states at the Fermi energy for channel $n$ of leads or islands $r$ and $r+1$, respectively. We define also the two-time coupling strength as $\alpha_0^{r} (t,t')=\sum_n N_r^n(0)\; N_{r+1}^n(0) \; V_{r} (t) V^*_r (t')$.

\section{Formalism}
We are interested in the dynamics of the excess charge on each island and not in the large number of noninteracting fermionic degrees of freedom which act effectively as baths.\cite{koenig96_1,koenig96_2,koenig_phd} Therefore, we trace over the states of the fermionic sector and we obtain the reduced density matrix $\varrho_{\text{red}}$, 
which just describes the charge degrees of freedom. In the basis defined by the states 
$|\chi\rangle=|N_1,\dots,N_M\rangle$ (with excess charge $N_r$ on the island $r$), the dynamics of the diagonal and off-diagonal matrix elements of the reduced density matrix are uncoupled. Furthermore, the off-diagonal matrix elements do not contribute to the current. 
Hence, we need only to consider the diagonal matrix elements, which are the occupation probabilities for the respective states and which we denote by $p_\chi (t)=\langle \chi|\varrho_{\text{red}}(t)|\chi\rangle$. 
The probabilities obey the generalized master equation
\begin{equation}
\label{eq_master}
\begin{split}
\dfrac{d}{dt}\mathbf{p}\left( t\right)
=\int\limits_{-\infty}^{t}dt'\;\mathbf{W}\left( t,t'\right)\mathbf{p}\left( t'\right) \,,
\end{split}
\end{equation}
where $\mathbf{p}$ is the vector of the probabilities $p_\chi$ and the matrix element $W_{\chi \chi'}(t,t')$ of the kernel $\mathbf{W}\left( t,t'\right)$ includes all transitions from state~$\chi'$ at time~$t'$ to state~$\chi$ at time~$t$.

\subsection{Adiabatic expansion}
We now proceed by performing an adiabatic expansion, i.e., a perturbation expansion in the frequency $\Omega$ of the time-dependent pumping fields $X_i(\tau)$. To describe the adiabatic pumping regime it is enough to retain terms linear in $\Omega$. 
The first step consists in performing  a Taylor expansion of $\mathbf{p}(t')$ on the right-hand side (r.h.s.) of Eq.~(\ref{eq_master}) around $t$ up to linear order,
\begin{equation}
	\frac{d}{dt}\mathbf{p}\left(t\right)=\int_{-\infty}^{t}dt'\mathbf{W}
	\left(t,t'\right)\left[\mathbf{p}\left(t\right)+(t'-t)\frac{d}{dt}
         \mathbf{p}(t)\right]\, .
\end{equation}
This expansion is justified if the period of the pumping fields is much longer than the characteristic memory time of the system. 
Second, we make an adiabatic expansion of the kernels appearing in the generalized master equation as follows:
\begin{equation}
\label{kernelexp}
  	\mathbf{W}(t,t') \rightarrow  \mathbf{W}^{(\text{i})}_{t}(t-t') + 
  	\mathbf{W}^{(\text{a})}_{t}(t-t'). 
\end{equation}
The instantaneous term, $\mathbf{W}^{(\text{i})}_{t}(t-t')$, is simply obtained 
by replacing $X_i(\tau)$  with $X_i(\tau=t)$. 
On the other hand, the first adiabatic correction, $\mathbf{W}^{(\text{a})}_{t}(t-t')$ is obtained by performing the Taylor expansion $X_i(\tau)\;\rightarrow\; 
	X_i(t)\;+\;\left(\tau-t\right)\;\left.\frac{d}{d\tau}X_i(\tau)\right|_{\tau=t}$ and retaining terms up to first order in the time derivatives.  
The index $t$ in Eq. (\ref{kernelexp}) indicates the time with respect to which the adiabatic expansion has been performed. 
The adiabatic expansion for the kernels is valid if the period of the pumping fields is much longer than the characteristic response times of the system, i.e., if $\Omega \ll \alpha_0 \max \left\lbrace \Delta_r, k_\text{B} T\right\rbrace $, where $\Delta_{r}$ are the charging-energy gaps of the islands. 

Finally, we expand also the probabilities,
\begin{equation}
  	\mathbf{p}(t)	\rightarrow  \mathbf{p}_t^{(\text{i})}\;+\;\mathbf{p}_t^{(\text{a})}	 \,.
\end{equation} 
Notice that by construction $\mathbf{p}_t^{(\text{i})}$ are the stationary probabilities for the time-independent problem where all parameters are frozen at their values at time $t$. 
Since all expressions now depend on the time difference $t-t'$ it is useful to introduce 
the Laplace transform $F\left(z\right)=\int_{-\infty}^{t}\;dt'\;e^{-z(t-t')}\;F(t-t')$. In the following we make use of the notation $\mathbf{W}_t^{(\text{i/a})}
	=\left.\mathbf{W}_t^{(\text{i/a})}\left(z\right)\right|_{z=0_+}
	=\int_{-\infty}^{t} dt'\,\mathbf{W}_t^{(\text{i/a})}(t-t')$ and 
$\partial \mathbf{W}_t^{(\text{i})}
	=\left.\frac{\partial}{\partial z}\; \mathbf{W}_t^{(\text{i})}\left(z\right)\right|_{z=0_+}$.

\subsection{Perturbative expansion in the tunnel coupling}
Even after having performed the adiabatic expansion, the problem cannot be solved exactly due to the presence of the Coulomb interaction and of the tunnel couplings. Therefore, in addition to the expansion in the pumping frequency $\Omega$, we perform a perturbative expansion in the 
tunnel-coupling strength $\alpha_0$. 
This limits the range of validity of the method to the weak tunnel-coupling regime.

The order in the expansion in $\alpha_0$ will be indicated by an integer superscript. For example, $\mathbf{W}^{(\text{a},1)}_t$ indicates the first order in $\alpha_0$ of the adiabatic correction of the kernel. 

The perturbative expansion for the instantaneous probabilities is straightforward since it corresponds to the time-independent problem with all parameters frozen at time $t$. 
The instantaneous probabilities $\mathbf{p}^{(\text{i},0)}_t$ and $\mathbf{p}^{(\text{i},1)}_t$ are determined from
\begin{eqnarray}
\label{instantan0}
  	0 & = & \mathbf{W}^{(\text{i},1)}_t \mathbf{p}^{(\text{i},0)}_t\\
\label{instantan1}
0 & = & \mathbf{W}^{(\text{i},2)}_t \mathbf{p}^{(\text{i},0)}_t 
  		+ \mathbf{W}^{(\text{i},1)}_t \mathbf{p}^{(\text{i},1)}_t, 
\end{eqnarray}
complemented by the normalization conditions $\mathbf{e}^{\text{T}} \mathbf{p}^{(\text{i},0)}_t =1$ and  $\mathbf{e}^{\text{T}} \mathbf{p}^{(\text{i},1)}_t =0$, where $\mathbf{e}$ is the  vector with all components equal to one. 
The first adiabatic correction to the probabilities obey the equations 
\begin{eqnarray}
\label{adiabatic-1}
  	\frac{d}{dt}\, \mathbf{p}^{(\text{i},0)}_t
		& = & \mathbf{W}^{(\text{i},1)}_t \mathbf{p}^{(\text{a},-1)}_t\\
\nonumber
\frac{d}{dt}\,\mathbf{p}^{(\text{i},1)}_t
		& = & \mathbf{W}^{(\text{i},1)}_t
  			\mathbf{p}^{(\text{a},0)}_t + \mathbf{W}^{(\text{i},2)}_t \mathbf{p}^{(\text{a},-1)}_t
\\
\label{adiabatic0}
&+&\mathbf{W}^{(\text{a},1)}_t\mathbf{p}^{(\text{i},0)}_t
			+ \partial \mathbf{W}^{(\text{i},1)}_t \frac{d}{dt}\,\mathbf{p}^{(\text{i},1)}_t,
\end{eqnarray}
complemented by the normalization conditions $\mathbf{e}^{\text{T}} \mathbf{p}^{(\text{a},-1)}_t =0$ and $\mathbf{e}^{\text{T}} \mathbf{p}^{(\text{a},0)}_t =0$. The orders in $\alpha_0$ on the r.h.s. of Eqs.~(\ref{adiabatic-1}) and (\ref{adiabatic0})  have been chosen to match those on the left-hand side. We notice that the expansion in $\alpha_0$ starts with the order 
$-1$. Although this result might seem surprising at first, it is in line with our expansion 
since  $\mathbf{p}^{(\text{a},-1)}_t \propto {\Omega}/ ( \alpha_0 \max \left\lbrace \Delta_r, k_\text{B} T\right\rbrace ) \ll 1$.

\subsection{Pumped current and charge}
The pumped current through the left lead can be written as
\begin{eqnarray}
	I_{\text{L}}\left(t\right)=e\int_{-\infty}^{t}dt'\mathbf{e}^{\text{T}}
	\mathbf{W}^{\text{L}}\left(t,t'\right)\mathbf{p}\left(t'\right)
\end{eqnarray}
with $e$ being the electron charge and $\mathbf{W}^{\text{L}}\left(t,t'\right)=\sum_p p
\mathbf{W}^{\text{L} p} \left(t,t'\right)$ . The kernel
$\mathbf{W}^{r p}\left(t,t'\right)$ describes all processes where the number of electrons entering reservoir $r$ minus the ones leaving it equals $p$. 

We proceed for the current in the same way as for the master equation~(\ref{eq_master}) performing an adiabatic expansion in the frequency $\Omega$ and a perturbative expansion in the tunnel coupling $\alpha_0$.
Since there is no applied transport voltage, the instantaneous current vanishes.
The adiabatic corrections to the current in zeroth and first order in $\alpha_0$ read
\begin{eqnarray}
\label{cur_ad_low}
  	I_{\text{L}}^{(\text{a},0)} (t) &=& e\ \mathbf{e}^{\text{T}}
  	\mathbf{W}^{\text{L}(\text{i},1)}_t \mathbf{p}^{(\text{a},-1)}_t\\
 \label{cur_ad_next}
 	I_{\text{L}}^{(\text{a},1)} (t) &=& e\ \mathbf{e}^{\text{T}} \left[
  	\mathbf{W}^{\text{L}(\text{i},1)}_t \mathbf{p}^{(\text{a},0)}_t
  	+ \mathbf{W}^{\text{L}(\text{i},2)}_t \mathbf{p}^{(\text{a},-1)}_t \right.
  	\nonumber \\
  	&&+  \left. \mathbf{W}^{\text{L}(\text{a},1)}_t \mathbf{p}^{(\text{i},0)}_t
  	+ \partial \mathbf{W}^{\text{L}(\text{i},1)}_t \frac{d\mathbf{p}_t^{(\text{i},0)}}{dt}
  	\right].
\end{eqnarray}

The lowest-order contribution to the adiabatic current, given by Eq.~(\ref{cur_ad_low}), scales with $\Omega$ and is independent of $\alpha_0$, in contrast to the dc current through a system with an applied transport voltage, which scales with $\alpha_0$. Since $\Omega \ll \alpha_0 \max \left\lbrace \Delta_r, k_\text{B} T\right\rbrace $, the pumped current goes to zero for vanishing tunnel coupling as it should. 
The pumped charge $Q$ per cycle is simply given by the integral of the pumped current over one period $\mathcal{T}=\frac{2 \pi}{\Omega}$.

\subsection{Diagrammatic rules}

\subsubsection{Rules to calculate the instantaneous kernels}

\begin{figure}
\begin{center}
\includegraphics[width=0.75 \columnwidth,clip]{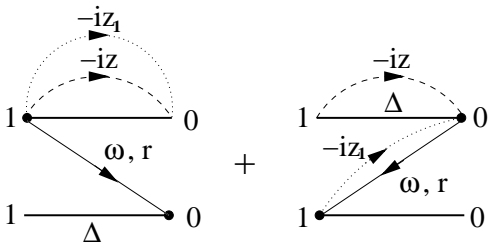}
\caption{Example of diagrams for the single-island system in first order in the tunnel coupling which are needed to calculate the kernels $\mathbf{W}_t^{(\text{i/a},1)}(z)$ and 
$\partial \mathbf{W}_t^{(\text{i},1)}(z)$. The additional frequency lines needed for the adiabatic corrections to the kernels are shown as dotted lines.
\label{fig1}}
\end{center}
\end{figure}

We start with summarizing the diagrammatic rules for the kernel 
$\mathbf{W}_t^{(\text{i},n)}(z)$, 
where $n$ indicates the order of the expansion in $\alpha_0$. Examples of first-order diagrams for the single-island system are drawn in Fig.~\ref{fig1}. 

\begin{itemize} 
\item[(1)] Draw all topologically different diagrams with $ n $ directed tunneling lines connecting pairs of vertices containing lead-electron operators. Assign a reservoir index $ r $ and energy $ \omega $ to each of these lines. Assign charge states $ \chi$ and the corresponding energies $ E_{\chi} (t) $ to each element of the Keldysh contour connecting two vertices. Furthermore, draw an external line from the upper leftmost beginning of an island propagator to the upper rightmost end of an island propagator that carries the (imaginary) energy $ -iz $.
\item[ (2)] For each time segment between two adjacent vertices (independent of whether they are on the same or on opposite branches of the Keldysh contour) write a resolvent $ \frac{1}{\Delta E(t)} $, where $ \Delta E(t) $ is the difference of left-going minus right-going energies (including energies of tunneling lines and the external line - the positive imaginary part of $ iz $ will keep all resolvents regularized).
\item[ (3)] Each vertex containing an island operator $ c_{knr}^{(\dagger)} $ gives rise to a matrix element $ \langle \chi'|c_{knr}^{(\dagger)}|\chi \rangle  $, where $ \chi\;\left( \chi'\right)  $ is the island state entering (leaving) the vertex with respect to the Keldysh contour.
\item[ (4)] The contribution of a tunneling line of reservoir $ r $ is $ \alpha_r^+(t,\omega)$ if the line is going backward with respect to the closed time path and \mbox{$ \alpha_r^-(t,\omega)$} if it is going forward with $\alpha_{r}^{\pm}(t,\omega)=\pm \alpha_{0}^{r}(t) \frac{\omega}{\exp(\pm \beta\omega)-1}$ and  $\beta=1/(k_\text{B} T)$.
\item[  (5)] The overall prefactor is given by $ (-i)(-1)^{b+c} $, where $ b $ is the total number of vertices on the backward propagator and $c$ the number of crossings of tunneling lines.
\item[  (6)] Integrate over all the energies of the tunneling lines and sum over the reservoir indices.
\end{itemize}

\subsubsection{Rules to calculate the adiabatic kernels}

The possible time-dependent parameters are the tunnel-matrix elements $V_{r}\left(t_i\right)$ and 
the charging-energy gaps $\Delta_{r}(t)$. 
The matrix elements $V_{r}\left(t_i\right)$ enter in the contractions of tunnel vertices as $\alpha_0^r\left(t_i,t_j\right) \propto  V_{r}\left(t_i\right)V^{*}_{r}\left(t_j\right)$. The charging-energy gaps $\Delta_{r}(t)$ enter in the isolated-island propagator  $\exp(- i\int_{t_i}^{t_j}dt' E_{\chi}\left(t'\right))$.

While for the instantaneous kernels all parameters were taken at time $t$, we now perform an expansion around the same time $t$ and keep all contributions linear in a time derivative of
the pumping parameters,
\begin{eqnarray}
	\alpha^{r}_0(t_i,t_j)&\approx& \alpha^{r}_0(t)\\
		&+&\;\dfrac{t_i-t}{2}\;\dfrac{d\alpha^{r}_0}{dt}(t)
		+\dfrac{t_j-t}{2}\;\dfrac{d\alpha^{r}_0}{dt}(t)    \nonumber\\
	e^{-i\int\limits_{t_i}^{t_j}dt'\;E_\chi (t')}&\approx& e^{-i\;E_\chi (t)\cdot(t_j-t_i)}
			\times \\
		&&\left[1-i\;\dfrac{(t_j-t)^2-(t_i-t)^2}{2}\;\dfrac{dE_\chi}{dt}(t) \right]\,. \nonumber
\end{eqnarray}
The linear factors $(t_i-t)$ appearing in these expansions can be included in the diagrammatic rules by introducing an additional external frequency line with 
the imaginary energy $-i z_i$ from the vertex at $t_i$ to the rightmost vertex 
at~$t$, performing the first derivative with respect to $z_i$, and then setting $z_i = 0_+$.
Similarly, the quadratic terms $(t_i-t)^2$ can be included in the diagrammatic rules by introducing an additional external frequency line with the imaginary energy $-i z_i$ from the beginning of an island-propagator line at~$t_i$ to the rightmost upper end of an island-propagator line at $t$, performing the second derivative with respect to $z_i$ and then setting $z_i = 0_+$.
 The external frequency lines are drawn as dotted lines in Fig.~\ref{fig1}.

The rules to  compute the contribution to the adiabatic corrections 
$\mathbf{W}_t^{(\text{a},n)}$ due to the time dependence of $\alpha^{r}_0(t)$ read:
\begin{itemize}
\item[ (7a) ]Add to all diagrams needed for $\mathbf{W}_t^{(\text{i},n)}(z)$ 
additional external frequency lines between any vertex $t_i$ and the upper right 
corner of the diagram and assign to them an (imaginary) energy $-iz_i$. 
Note that an external frequency line between two right corners of a 
diagram does not contribute and can always be omitted.
\item[ (7b)] Follow the rules (1) to (6) taking into account the extra lines.
\item[ (7c)] Perform a first derivative with respect to $ z_i $ and multiply it by $ \frac{1}{2} \frac{d\alpha^{r}_0}{dt}(t) \frac{1}{\alpha^{r}_0(t)}$. Sum all the contributions obtained in this way.
\item[ (7d)] Set all the external frequencies $z_i$ and $z$ to $0_+$.
\end{itemize}

The contribution to the adiabatic correction  $\mathbf{W}_t^{(\text{a},n)}$ due to 
the  time dependence of the charging energy can be computed in a similar way:  
\begin{itemize}
\item[ (8a)] In addition to the external frequency lines added according to rule (7a), 
put one more external frequency line from the left corner of the diagram 
with no vertex to the right corner.  
\item[ (8b)] Follow the rules (1) to (6) taking into account the extra lines.
\item[ (8c)] Perform a second derivative with respect to $ z_i $ and multiply it by $ -\frac{i}{2} \frac{d(E_{\chi}-E_{\chi'})}{dt}(t) $, where $ \chi\;(\chi') $ is the island state entering (leaving) the vertex of the external frequency line at $ t_i $ with respect to the Keldysh contour. The term $ \frac{dE_\chi}{dt} \; \left( \frac{dE_{\chi'}}{dt}\right) $ is omitted if the segment associated with $ E_{\chi}\; \left(E_{\chi'} \right)  $ does not belong to the diagram. Sum all the contributions obtained in this way. 
\item[ (8d)] Set the external frequencies $z_i$ and $z$ to $0_+$.
\end{itemize}

Finally to compute the current, we need the matrix elements of $\mathbf{W}_t^r=\sum_p\; p\;\mathbf{W}^{r p}_t$. For each contribution $\mathbf{W}^{rp}_t$ all diagrams for which the number of tunneling lines with reservoir index $ r $ running from the upper to the lower propagator minus the number of those with reservoir index $ r $ running in the opposite direction equals $ p $ need to be added.

\section{Adiabatic Pumping through a Single-Island System}

In this section, we present the results for a  single-island system consisting of one metallic island with Coulomb interaction, tunnel coupled to two noninteracting leads kept at the same chemical potential. 
The Coulomb interaction on the island is described by the charging energy $
E_{\text{ch}}(N,N_{x}(t)) =   E_{\text{C}} (N-N_{x}(t))^2$,
where $N_{x}(t)$ denotes the charges externally induced via the gate voltage $V_{\text{G}}$ and $N$ indicates the charge on the island. The energy scale for the charging energy is $ E_{\text{C}} =\frac{e^2}{2C}$ with $C$ being the total island capacitance. 

\begin{figure}
\begin{center}
\includegraphics[height=0.35 \columnwidth,clip]{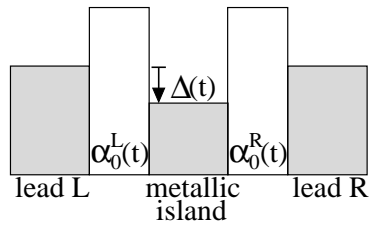}
\caption{Sketch of the single-island system. 
The possible pumping parameters are  the charging-energy gap $\Delta$ of the island and the tunnel-coupling strengths $\alpha^{r}_{0}$.
\label{fig2}}
\end{center}
\end{figure}

At low temperatures only two different charge states contribute to the transport. 
We indicate these states as $|0\rangle$ and $|1\rangle$ and their  energies as $E_0$ and $E_{1}$, respectively. 
The occupation  probabilities for these states are labeled   $p_0$ and $p_1$. 
The charging-energy gap of the island is defined as the energy difference $ \Delta(t) \equiv E_1(t) - E_0(t)$\,.
An energy sketch of the single-island system is shown in Fig.~\ref{fig2}.

\subsection{Pumped current in zeroth order in $\alpha_0$}

To calculate the pumped current in zeroth order in $\alpha_0$ we need the instantaneous occupation probabilities in zeroth order in $\alpha_0$  and their adiabatic correction in minus first order. 
The instantaneous probabilities  can be obtained from Eq.~(\ref{instantan0}) and are simply 
given by 
Boltzmann factors,
\begin{equation}
\label{eq_p_i,0}
\left( \mathbf{p}_t^{(\text{i},0)}\right)_{\chi} =\frac{e^{-\beta E_{\chi}}}{Z} \, ,
\end{equation}
where $\beta=\frac{1}{k_{\text{B}}T}$ is the inverse temperature, $E_{\chi}$ the energy of the island state 
$\chi$, and $Z=\sum_\chi e^{-\beta E_{\chi}}$ the partition function. 
The instantaneous average occupation number of the island in lowest order in $\alpha_0$ can be calculated from the instantaneous probabilities  and  reads
\begin{equation}
\label{eq_n_i,0}
\left\langle n \right\rangle ^{(\text{i},0)} 
= f^+(\Delta)  \, .
\end{equation}

The adiabatic correction to the probabilities can be obtained from Eqs.~(\ref{adiabatic-1}) and (\ref{eq_p_i,0})  and  is given by
\begin{equation}
\label{ad_prob_-1}
\mathbf{p}_t^{(\text{a},-1)}=-\;\dfrac{1}{2 \pi \,\alpha(\Delta)}\quad\dfrac{d}{dt} \mathbf{p}_t^{(\text{i},0)}\,,
\end{equation}
where we used the notation  $\alpha(\omega)\;=\alpha^+(\omega) +\alpha^-(\omega)$ and $\alpha^\pm(\omega)=\sum_r\alpha_r^\pm(\omega)$. Furthermore, we define $\alpha_0 =\alpha_0^\text{L} +\alpha_0^\text{R}$.
Since this probability vector is proportional to the time derivative of the instantaneous probabilities $\mathbf{p}^{(\text{i},0)}_t$, it is an eigenvector of the kernel $\mathbf{W}^{(\text{i},1)}_t$. 
To compute the probabilities Eqs. (\ref{eq_p_i,0}) and (\ref{ad_prob_-1}), we need to evaluate only the  instantaneous kernels in first order in $\alpha_0$. The expressions for the matrix elements of $\mathbf{W}^{(\text{i},1)}_t$ are given in  Appendix~\ref{appendix_A} .

By inserting the results for the probabilities in Eq.~(\ref{cur_ad_low}), it yields for the adiabatically pumped current through the single-island system in zeroth order $\alpha_0$,
\begin{equation}
\label{cur_a_0}
	I_\text{L}^{(\text{a},0)}(t)=-\;e\;\dfrac{\alpha_0^\text{L}}{\alpha_0}\; 
		\;\dfrac{d}{dt}\, \left\langle n \right\rangle ^{(\text{i},0)}\, .
\end{equation}
The average occupation number of the island in zeroth order $\alpha_0$ depends  only on the charging-energy gap $\Delta$ and not on the tunnel-coupling strengths. Therefore,  
 the pumped current in zeroth order $\alpha_0$ is nonvanishing only if the 
charging-energy gap is one of the pumping parameters.
The result of Eq.~(\ref{cur_a_0}) has a simple interpretation:  if the charging-energy gap is varied in time, the occupation of the island adjusts to the new situation and therefore electrons need to enter or leave the island. The fraction of the resulting electrical current which flows through barrier
$r$ is given  by the ratio $\alpha_0^r/\alpha_0$. Therefore, this pumping mechanism can be considered as  peristaltic.

\subsection{Pumped current in first order in $\alpha_0$}
\label{sec}
We start by giving the expression for the instantaneous average occupation number of the island in first order in $\alpha_0$ as follows:
\begin{equation}
\label{eq_n_i,1}
\begin{split}
\left\langle n \right\rangle ^{(\text{i},1)} 
= \; \alpha_0 \;&\biggl\lbrace 
\quad \Re\left[ \varrho(\Delta)\right] \;\dfrac{d}{d \Delta} \left\langle n\right\rangle ^{(\text{i},0)}\biggr.\\
&\;+\biggl.\Bigl( \left\langle n\right\rangle ^{(\text{i},0)}\;-\;\dfrac{1}{2}\Bigr) \;\;
\partial_\Delta\;\Re\left[ \varrho(\Delta)\right]
\biggr\rbrace  \,, 
\end{split}
\end{equation}
The function $\varrho(\omega)$ is defined via   $\varrho(\omega)=\varrho^+(\omega) + \varrho^-(\omega)$ with $\varrho^\pm(\omega)=\pm\int d\omega'\;\frac{\omega'}{\omega-\omega'+i0_+}\; \frac{1}{e^{\pm\beta \omega'}-1}$. To keep the integrals convergent we introduce a  high-energy cutoff, which in our case is provided by the charging-energy scale $E_\text{C}$. Then in the limit of $E_\text{C} \gg {\rm max} \lbrace \omega, k_\text{B} T \rbrace$ we obtain for the real part of the $\varrho$-function,
\begin{equation}
	\Re\left[ \varrho(\omega) \right]
		=- 2\; \omega   \;
		\Biggl\lbrace  \ln \left( \dfrac{\beta E_\text{C}}{2\pi } \right)
		-  \Re\psi\left( i \dfrac{\beta \omega}{2\pi} \right)\Biggr\rbrace \,,
\end{equation}
where $\psi$ denotes the digamma function. 
We identify the first term in Eq.~(\ref{eq_n_i,1}) as the contribution due to the renormalization of the charging-energy gap $\Delta$,
\begin{equation}
	\left\langle n\right\rangle ^{(\text{i},\text{ren})}
		=\alpha_0 \;\Re\left[\varrho(\Delta)\right]\; \dfrac{d}{d 
		\Delta} \left\langle n\right\rangle ^{(\text{i},0)} \,,
\end{equation}
where the charging-energy gap is renormalized by $\Delta \rightarrow \Delta \,+\, \alpha_0 \;\Re\left[\varrho(\Delta)\right]$.  

The broadening of the spectral function of the island due to the tunnel coupling to the leads yields a contribution to the first-order correction of the island occupation number, which can be written as the sum of the contributions of the individual barriers, $\left\langle n\right\rangle ^{(\text{i},\text{broad})}= \left\langle n \right\rangle ^{(\text{i},\text{broad},\text{L})} \;+\; \left\langle n \right\rangle^{(\text{i},\text{broad},\text{R})}$.
The second term of Eq.~(\ref{eq_n_i,1}) can be identified as the contribution due to broadening,
\begin{equation}
	\left\langle n\right\rangle ^{(\text{i},\text{broad})}
	=\Bigl( \left\langle n\right\rangle ^{(\text{i},0)}\;-\;\dfrac{1}{2}\Bigr) 
		\; \alpha_0 \; \partial_\Delta\;\Re\left[\varrho(\Delta)\right] \,,
\end{equation}
where $\partial_\Delta$ denotes the derivative with respect to the charging-energy gap $\Delta$. The amplitude of the broadening depends on the occupation of the island: if the island is unoccupied, the prefactor is $-\frac{1}{2}$; if the island is occupied by a single electron, the coefficient is $\frac{1}{2}$.

In order to compute the first-order correction to the pumped current, we need, besides $\mathbf{p}^{(\text{i},1)}_t$, also $\mathbf{p}^{(\text{a},0)}_t$, which can be computed by means of Eq.~(\ref{adiabatic0}) and reads
\begin{equation}
\label{p_a,0}
\begin{split}
\left( \mathbf{p}_t^{(\text{a},0)}\right)_{1} &=\;-\left( \mathbf{p}_t^{(\text{a},0)}\right)_{0} \\
&=\quad\dfrac{1}{\alpha^2(\Delta)}\;\dfrac{\alpha_0}{2\pi}
 \;   \Re\left[ \varrho(\Delta)\right] \; \partial_\Delta\alpha(\Delta)
\;\;\dfrac{d}{dt}\,\left[ f^+(\Delta)\right]\\
&\quad
\; -\;\dfrac{1}{\alpha(\Delta)}\;
\dfrac{d}{dt}\left( 
 \dfrac{\alpha_0}{2\pi}\;\;
\Re\left[ \varrho(\Delta)\right]\;\partial_\Delta f^+(\Delta)\; 
\right) \;.
\end{split}
\end{equation}
The expressions of the matrix elements of kernels  $\mathbf{W}^{(\text{a},1)}_t$ and $\mathbf{W}^{(\text{i},2)}_t$, which are needed to compute the probabilities Eq.~(\ref{p_a,0}), 
can be found in Appendix~\ref{appendix_A}. Finally, 
the pumped current in first order in $\alpha_0$ can be written as
\begin{equation}
\label{cur_a_1}
	I_\text{L}^{(\text{a},1)}(t)=-e
\left\lbrace \dfrac{d}{dt} \left\langle n\right\rangle ^{(\text{i},\text{broad},\text{L})}
+\dfrac{\alpha_0^\text{L}}{\alpha_0}\; \dfrac{d}{dt}\left\langle n\right\rangle ^{(\text{i},\text{ren})}\right\rbrace\,.
\end{equation}
The first term is related to the time variation of the correction of the average occupation number due to broadening of the resonance at $\Delta$ induced by tunnel coupling the island to the left lead. It is, therefore, associated with tunneling processes where electrons tunnel through the left barrier. Equation~(\ref{cur_a_1}) shows that this first contribution to the current consists of the total time derivative of the correction of the average occupation number due to tunnel coupling to the left lead and thus the integral over a whole pumping period yields zero. 
The second contribution to the pumped current in first order in $\alpha_0$ is due to the renormalization of the charging-energy gap and has the same form as the zeroth-order contribution to the current, Eq.~(\ref{cur_a_0}), with renormalized parameters. As the renormalized average occupation number of the island $\left\langle n \right\rangle ^{(\text{i},\text{ren})}$ depends on the charging-energy gap as well as on the tunnel-coupling strengths $\alpha_0^r$, we obtain a nonvanishing pumped charge  for any choice  of pumping parameters. Particularly intriguing is the case of pumping with the 
two barriers, since in this case the current $I_\text{L}^{(\text{a},0)}(t)$ vanishes and the 
first nonvanishing contribution to the current is due to the renormalization of the charging-energy gap, and therefore it is an effect purely due to quantum fluctuations and interaction.

\subsection{Pumped charge in the weak-pumping limit}

Now we calculate the pumped charge per cycle in the weak-pumping limit. We write the pumping parameters 
as  $X_i(t)=\bar{X_i} +\delta X_i(t)$, where $\bar{X_i}$ is the time average and $\delta X_i(t)$ the oscillating part.
The weak-pumping limit  corresponds to computing 
the current in  bilinear order in the time-varying parts of the parameters.
 
\subsubsection{Pumping with the charging-energy gap and one tunnel-coupling strength}

We start by considering  one tunnel-coupling strength $\alpha_0^\text{L}$ and the charging-energy gap $\Delta$ as pumping parameters. Expanding the current  $I_\text{L}^{(\text{a})}(t)$ up to bilinear order in the time variation of the pumping parameters and integrating over one cycle, we obtain for the pumped charge
\begin{equation}
\begin{split}
Q_{\alpha_0^\text{L}, \Delta}^{(\text{a})} &=-\; e \;\dfrac{ \bar{\alpha}_0^\text{R} } {\bar{\alpha}_0 ^2 }\beta \; 
	\eta_1\times\\
&\dfrac{d}{d\beta\bar{\Delta}} \left\lbrace \left\langle \bar{n}\right\rangle ^{(\text{i},0)}  +\; \bar{\alpha}_0\beta \; \Re\left[\varrho({\bar{\Delta}})\right]\; \dfrac{d}{d\beta\bar{\Delta}} \left\langle \bar{n}\right\rangle ^{(\text{i},0)} \right\rbrace \,,
\end{split}
\end{equation}
where $\eta_1 =\int _0^{\mathcal{T}}\; \delta\alpha_0^\text{L}\; \frac{d\;\delta\Delta}{dt}\;  dt$ is the area of the pumping cycle in parameter space, $\left\langle \bar{n}\right\rangle ^{(\text{i},0)}$ the instantaneous average occupation number, where the charging-energy gap has been replaced with its time average ${\bar{\Delta}}$, and $\bar{\alpha}_0 =\bar{\alpha}_0^\text{L} +\bar{\alpha}_0^\text{R}$.

The first contribution to the pumped charge arises from the current in zeroth order in $\alpha_0$, the second from the first-order correction of the current due to the renormalization of the charging-energy gap as discussed in Sec.~\ref{sec}. Hence, the first term is the dominant one.

\begin{figure}
\begin{center}
\includegraphics[width=\columnwidth,clip]{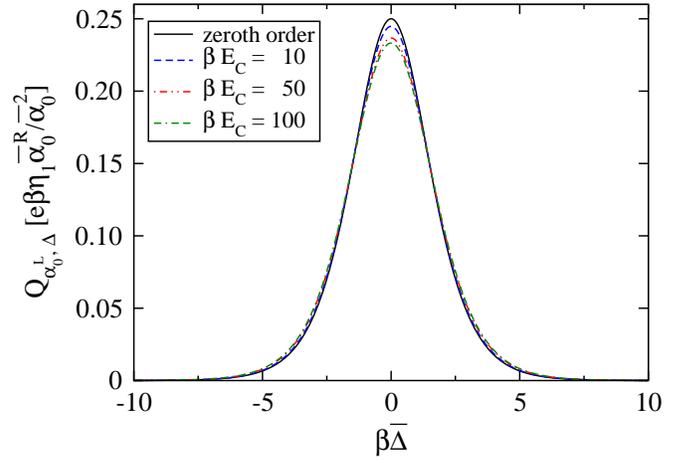}
\caption{(Color online)
Pumped charge through the single-island system up to first order in $\alpha_0$ in units of $e 
 \beta \eta_1 \bar{\alpha}_0^\text{R}/ {\bar{\alpha}_0 ^2 } $ as a function of the time average of the charging-energy gap for different values of $E_\text{C}$. The pumping parameters are $\alpha_0^{\text{L}}$ and $\Delta$. The tunnel-coupling strength is $\bar{\alpha}_0=0.01$.
\label{fig3}}
\end{center}
\end{figure}

In Fig.~\ref{fig3} the pumped charge is plotted as a function of the time average of the charging-energy gap for different values of the high-energy cutoff. The pumped charge is symmetric around zero and has its maximum there. This peak is associated with transitions between charge states where the island is unoccupied or occupied by a single electron. The pumped charge vanishes for charging energies much bigger than the temperature ($|{\bar{\Delta}}|\gg k_\text{B} T$) without changing its sign. The overall sign of the pumped charge is determined by the traversal direction of the pumping cycle in parameter space. In Fig.~\ref{fig3} we can clearly see that the maximum of the pumped charge decreases for increasing high-energy cutoff due to  the correction introduced by the renormalization of the charging-energy gap.

\begin{figure}
\begin{center}
\includegraphics[width=\columnwidth,clip]{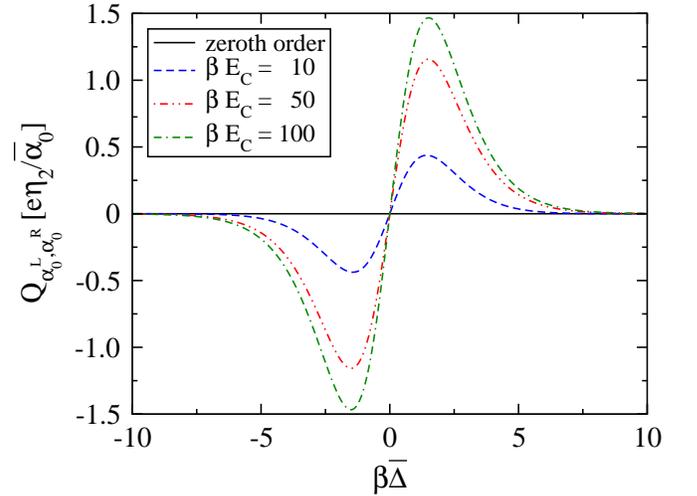}
\caption{(Color online)
Pumped charge through the single-island system up to  first order in $\alpha_0$ in units of $e\eta_2/\bar{\alpha}_0$ as a function of the time average of the charging-energy gap  for different values of $E_\text{C}$. The pumping parameters are $\alpha_0^{\text{L}}$ and $\alpha_0^{\text{R}}$. 
\label{fig4}}
\end{center}
\end{figure}

\subsubsection{Pumping with the tunnel-coupling strengths}

When we choose both tunnel-coupling strengths $\alpha_0^\text{L}$ and $\alpha_0^\text{R}$ as pumping parameters the contribution in zeroth order in $\alpha_0$ vanishes, as already mentioned. We therefore obtain for the pumped charge in the weak-pumping limit only the contribution due to the renormalization of the charging-energy gap,
\begin{equation}
	Q_{\alpha_0^\text{L}, \alpha_0^\text{R}}^{(\text{a})}=\; e \; \dfrac{\eta_2 } {\bar{\alpha}_0 ^2 }\; 
		 \bar{\alpha}_0   \;\Re\left[ \varrho({\bar{\Delta}})\right]\;\dfrac{d}{d\bar{\Delta}} \left\langle \bar{n}\right\rangle ^{(\text{i},0)}\,,
\end{equation}
where $\eta_2 =\int_0^{\mathcal{T}} \; \delta\alpha_0^\text{R} \; \frac{d\;\delta\alpha_0^\text{L}}{dt}  \; dt$ is again the area of the pumping cycle in parameter space. The pumped charge  as a function of the time-averaged charging-energy gap is plotted in Fig.~\ref{fig4}. It is zero for $\bar{\Delta}=0$ and  antisymmetric around zero.  For charging energies much larger than the temperature ($|{\bar{\Delta}}|\gg k_\text{B} T$) the pumped charge vanishes. Increasing the high-energy cutoff $E_\text{C}$, the amplitude of the pumped charge is increased. Finally, we emphasize that by adiabatic pumping with the two tunnel-coupling strengths we can explore the effects of renormalization of the charging-energy gap $\Delta$. The pumping mechanism is peristaltic but it occurs via renormalization effects. The sign change of the pumped charge around the resonance clearly distinguishes this mechanism from lowest-order pumping with one barrier and gate voltage. We remark that such a sign change has been experimentally observed in charge pumping through carbon nanotubes with the help of surface acoustic waves.\cite{buitelaar} There, this feature has been understood as adiabatic pumping through an effective few-level double quantum dot with changing level positions.

\section{Adiabatic Pumping through a Double-Island System}

\begin{figure}
\begin{center}
\includegraphics[height=0.35 \columnwidth,clip]{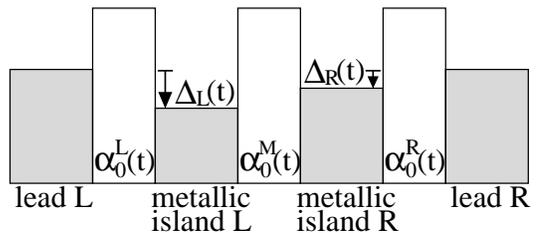}
\caption{Sketch of the double-island system.   It consists of two metallic islands tunnel coupled in series to noninteracting leads which are kept at the same chemical potential. We consider as  pumping parameters the two charging-energy gaps $\Delta_{r}$. 
\label{fig5}}
\end{center}
\end{figure}

In this section we consider pumping with the two charging-energy gaps in a system consisting of two tunnel-coupled metallic islands with no applied bias voltage. This system corresponds to the single-electron pump of Pothier {\it et al.} \cite{pothier92} 
The left and right islands are labeled by $\text{L}$ and $\text{R}$, respectively. 
Notice that we will use the same labels for the the left and right leads. Whenever confusion could arise, we specify whether we refer to an island or a lead. 
The charging energy of the system reads
\begin{equation}
\begin{split}
&E_{\text{ch}} (N_\text{L}, N_\text{R}, N_{x,\text{L}}(t), N_{x,\text{R}}(t))  \\
=&\quad   E_{\text{CL}} (N_\text{L}-N_{x,\text{L}})^2  +  E_{\text{CR}} (N_\text{R}-N_{x,\text{R}})^2 \\
& +   E_{\text{CM}} (N_\text{L}-N_{x,\text{L}}) (N_\text{R}-N_{x,\text{R}})\,,
\end{split}		
\end{equation}
where $N_{\text{L}(\text{R})}$ is the total charge on the left (right) island and  
$ N_{x,\text{L}(\text{R})}$ the gate-induced charge on the left (right) island.
The prefactors $E_{\text{CL}}$, $E_{\text{CR}}$, and $E_{\text{CM}}$ can be easily computed as a function of the junction and gate capacitances. 
At low temperature, the Hilbert space for the double-island system can be truncated to the three states 
$|0,0\rangle$,  $|1,0\rangle$, and  $|0,1\rangle$, corresponding to no excess charge in the system, one excess charge on the left island and none on the right one, and one excess charge on the right island and none on the left one, respectively. The corresponding energies are
$E_{0,0}(t)$, $E_{1,0}(t)$, and $E_{0,1}(t)$. 
We define the two time-dependent charging-energy gaps $ \Delta_\text{L}(t) \equiv E_{1,0}(t) - E_{0,0}(t)$ and $ \Delta_\text{R}(t) \equiv E_{0,1}(t) - E_{0,0}(t)$ of the islands and the abbreviation $ \Delta_\text{M}(t) \equiv \Delta_\text{R}(t) - \Delta_\text{L}(t)$. 
The occupation probabilities for these charge states are $p_0$, $p_\text{L}$, and $p_\text{R}$ and 
are collected in the  probability vector $\mathbf{p}=\left(p_0,p_\text{L},p_\text{R}\right)^{\text{T}}$ .
A sketch of the double-island system is shown in Fig.~\ref{fig5}.

In comparison to the single-island system, the tunneling events in the double-island system include additional tunneling between the two islands. We define the direction of these 
tunneling lines to correspond to an electron tunneling from the right to the left island. 
Since no transport voltage is applied the pumped current vanishes in instantaneous order and we need to compute the first adiabatic correction. As in this system the current in zeroth order in the tunnel-coupling strength is the dominant term, we restrict ourselves to this contribution.

\subsection{Pumped current in zeroth-order in $\alpha_0$}

To compute the pumped current in zeroth order in $\alpha_0$ we need the probabilities $\mathbf{p}^{(\text{a},-1)}_t$, determined by Eq.~(\ref{adiabatic-1}), together with Eq.~(\ref{instantan0}). For this purpose, only the instantaneous kernels in first order in $\alpha_0$ are necessary. 
Examples of diagrams contributing to the matrix elements of the kernel $\mathbf{W}^{(\text{i},1)}_t$ for the double-island system can be found in Appendix~\ref{appendix_B}. 
Following this procedure we find for the instantaneous occupation probabilities
\renewcommand{\arraystretch}{1.25}
\begin{equation}
\label{prob-i-0}
\mathbf{p}_t^{(\text{i},0)}=
\frac{1}{e^{-\beta  \Delta_\text{L}} + e^{-\beta  \Delta_\text{R}} + 1}
\left( 
\begin{array}{c}
1\\
e^{-\beta \Delta_\text{L} }\\
e^{-\beta \Delta_\text{R} }
\end{array}
\right)\,.
\end{equation}
Making use of the notation $\alpha_{r}^{\pm}=\alpha_{r}^{\pm}(\Delta_s)$ and $\alpha_r =\alpha_r^+ + \alpha_r^-$ with $r\in\left\lbrace \text{L},\text{R},\text{M} \right\rbrace$ as well as
$\mathcal{N} =\alpha_\text{L} \alpha_\text{M}^- + \alpha_\text{L}^+ \alpha_\text{M}^+ + \alpha_\text{L} \alpha_\text{R}^- + \alpha_\text{L}^-  \alpha_\text{R}^+  + \alpha_\text{M}  \alpha_\text{R}^+  + \alpha_\text{M}^+ \alpha_\text{R}^-$ for the denominator, the adiabatic probabilities can be written as
\begin{equation}
\begin{split}
\mathbf{p}_t^{(\text{a},-1)}
&=\dfrac{1}{2 \pi \, \mathcal{N}} 
\left( 
\begin{array}{ccc}
0 & \alpha_\text{M}  + \alpha_\text{R}^- & \alpha_\text{M} + \alpha_\text{L}^- \\
\alpha_\text{R}  + \alpha_\text{M}^-  & 0 &  \alpha_\text{R} + \alpha_\text{L}^+ \\
\alpha_\text{L} + \alpha_\text{M}^+ & \alpha_\text{L} + \alpha_\text{R}^+ & 0
\end{array}
\right) \dfrac{d}{dt} \; \mathbf{p}_t^{(\text{i},0)} \,.
\end{split}
\end{equation}

By substituting these results in Eq.~(\ref{cur_ad_low}) we obtain the adiabatically pumped current in the left lead  in zeroth order in $\alpha_0$,
\begin{equation}
\label{cur_exp}
	I_\text{L}^{(\text{a},0)}(t) =e \left\lbrace g_\text{R} \, \dfrac{d}{dt} \left(
	 \mathbf{p}_t^{(\text{i},0)}\right)_\text{R} +g_\text{L} \, \dfrac{d}{dt} \left(
	 \mathbf{p}_t^{(\text{i},0)}\right)_\text{L}  \right\rbrace \,,
\end{equation}
where the functions $g_\text{L}$ and $g_\text{R}$ are defined by
\begin{subequations}
\begin{eqnarray}
g_\text{R}&=&-\;\dfrac {\alpha_\text{L} \alpha_\text{M}^- + \alpha_\text{L}^+ \alpha_\text{M}^+ }{\mathcal{N}}\\
g_\text{L}&=&-\;\dfrac{\alpha_\text{L} \alpha_\text{M}^- + \alpha_\text{L}^+ \alpha_\text{M}^+ + \alpha_\text{L} \alpha_\text{R}^- + \alpha_\text{L}^-  \alpha_\text{R}^+ }{\mathcal{N}} \,.
\end{eqnarray}
\end{subequations}
The current, Eq.~(\ref{cur_exp}), depends on the time derivative of the occupation probabilities of both islands.
The current in the right lead can be obtained by exchanging L and R in Eq.~(\ref{cur_exp}); in doing so we have to replace also $\alpha_\text{M}^\pm$ with $\alpha_\text{M}^\mp$.

\subsection{Pumped charge}

To compute the pumped charge per cycle we simply need to integrate  the current $I_\text{L}^{(\text{a},0)}(t)$, given in Eq.~(\ref{cur_exp}),  over one period. 
Using simple mathematical manipulations, the pumped charge per period can be cast in the form
\begin{equation}
Q_{\Delta_\text{L},\Delta_\text{R}}=e \int_0^\mathcal{T} dt  \sum\limits_{j=\left\lbrace \text{R},\text{L}\right\rbrace}
 \frac{d D_\text{L}}{d \Delta_{j}} \frac{d \Delta_j}{dt},
\end{equation}
where the emissivity for the left lead is given by 
\begin{equation}
 \frac{d D_\text{L}}{d \Delta_{j}}= g_\text{R} \, \dfrac{d}{d \Delta_j} \left(
	 \mathbf{p}_t^{(\text{i},0)}\right)_\text{R} +g_\text{L} \, \dfrac{d}{d\Delta_j} \left(
	 \mathbf{p}_t^{(\text{i},0)}\right)_\text{L}.
\end{equation}
 Following Ref.~\onlinecite{brouwer98} the pumped charge per period can be written as an integral over the area $A$ spanned by the cycle in parameter space and reads 
\begin{equation}
\label{brouwer}
Q_{\Delta_\text{L},\Delta_\text{R}}=e \int_A d   \Delta_{\text{R}} d   \Delta_{\text{L}} \left\{ \frac{\partial} {\partial\Delta_{\text{R}}} 
 \frac{d D_{\text{L}}}{d \Delta_{\text{L}}} -  \frac{\partial}{\partial \Delta_{\text{L}}} 
 \frac{d D_{\text{L}}}{d \Delta_{\text{R}}} \right\} .
\end{equation}
We now consider the weak-pumping regime and we write the pumping parameters, here the charging-energy gaps, as their mean values plus a small time-dependent variation around it, i.e., $\Delta_r(t)=\bar{\Delta}_r +\delta\Delta_r(t)$.

Equation~(\ref{brouwer}) in the weak-pumping limit yields for the zeroth-order contribution to the pumped charge
\begin{equation}
\label{charge_exp}
\begin{split}	
Q_{\Delta_\text{L},\Delta_\text{R}}
	=e\,\eta_3  \sum\limits_{j=\left\lbrace \text{R},\text{L}\right\rbrace } \, 
	& \left[\quad
	\dfrac{\partial \,\bar{g}_j}{\partial\,{\bar{\Delta}}_\text{R}} \; 
	\dfrac{\partial}{\partial\,{\bar{\Delta}}_\text{L}} \left( \mathbf{p}_t^{(\text{i},0)}\right)_{j}\right.\\
	&\;-\;\left.
	\dfrac{\partial \,\bar{g}_j}{\partial\,{\bar{\Delta}}_\text{L}} \; 
	\dfrac{\partial}{\partial\,{\bar{\Delta}}_\text{R}}  \left( \mathbf{p}_t^{(\text{i},0)}\right)_{j} 
	\right] \,,
\end{split}
\end{equation}
where $\eta_3=\int_0^{\mathcal{T}} \; \delta\Delta_\text{R} \; \frac{d\;\delta\Delta_\text{L}}{dt}  \; dt$ is the area enclosed by the pumping cycle in parameter space. In the functions $\bar{g}_j$ and the instantaneous occupation probabilities $( \mathbf{p}_t^{(\text{i},0)})_{j}$ the charging-energy gaps have been replaced with their time averages ${\bar{\Delta}_j}$.

\begin{figure}
\begin{center}
\includegraphics[width=\columnwidth,clip]{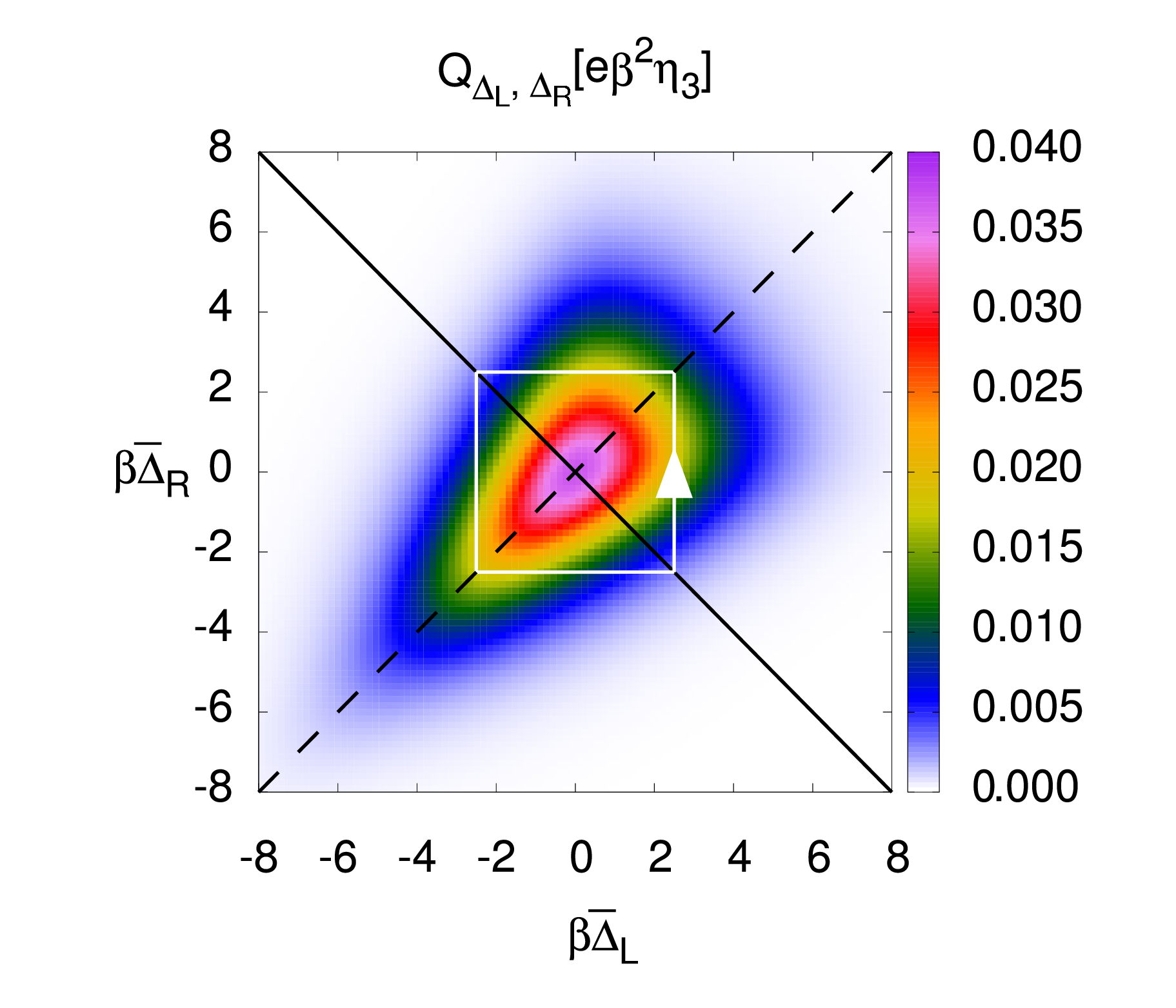}
\caption{(Color online)
Pumped charge through the double-island system  in the weak-pumping limit in zeroth order in $\alpha_0$ in units of $e\beta^2\eta_3$ as a function of the time averages of the pumping parameters ${\bar{\Delta}}_\text{L}$ and ${\bar{\Delta}}_\text{R}$  in units of the inverse temperature $\beta^{-1}=k_\text{B} T$. 
The white square marks the area in parameter space over which we integrate to obtain the pumped charge in the strong-pumping limit.
\label{fig6}}
\end{center}
\end{figure}
\begin{figure}
\begin{center}
\includegraphics[width=1.\columnwidth,clip]{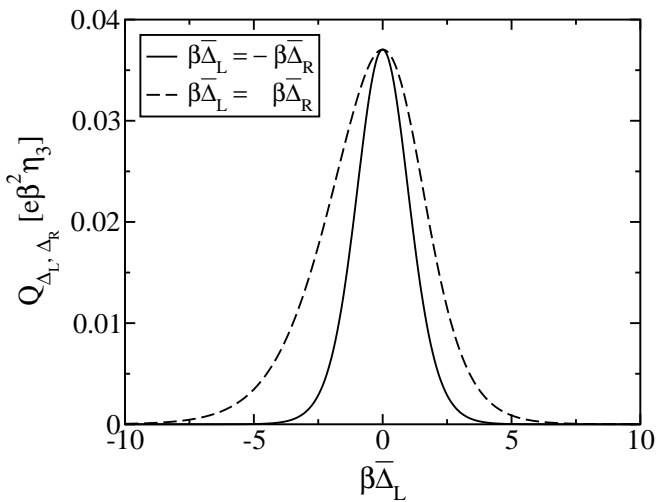}
\caption{Pumped charge through the double-island system  in the weak-pumping limit in zeroth order in $\alpha_0$ in units of $e\beta^2\eta_3$  for $\Delta_\text{L}$ equal to $\Delta_\text{R}$ (dashed line) and  $\Delta_\text{L}$ equal to $-\Delta_\text{R}$ (solid line).
\label{fig7}}
\end{center}
\end{figure}

A density plot of the pumped charge as a function of the time averages of the pumping parameters ${\bar{\Delta}}_\text{L}$ and ${\bar{\Delta}}_\text{R}$  is shown in Fig.~\ref{fig6}. It consists in a peak with its maximum in the origin (for symmetrically chosen tunnel couplings $\alpha_0^\text{L}=\alpha_0^\text{R}=\alpha_0^\text{M}$). This can be seen more clearly in Fig.~\ref{fig7} where we show a cross section of the peak along the line with ${\bar{\Delta}}_\text{L}=-{\bar{\Delta}}_\text{R}$ (solid line) and along the line with ${\bar{\Delta}}_\text{L}={\bar{\Delta}}_\text{R}$ (dashed line).  Notice that the peak is symmetric for the case of 
antisymmetrically  chosen  charging-energy gaps, while the symmetry is absent for symmetrically chosen charging-energy gaps.
Going around the origin of the plot in Fig.~\ref{fig6} with constant radius $R=\sqrt{{\bar{\Delta}}_\text{L}^2\;+\;{\bar{\Delta}}_\text{R}^2}$, the pumped charge exhibits a maximum when the polar angle takes the values $\frac{\pi}{4}$ and 
 $\frac{5}{4}\pi$,  which correspond to ${\bar{\Delta}}_\text{L}={\bar{\Delta}}_\text{R}$\,.

\begin{figure}
\begin{center}
\includegraphics[width=1.\columnwidth,clip]{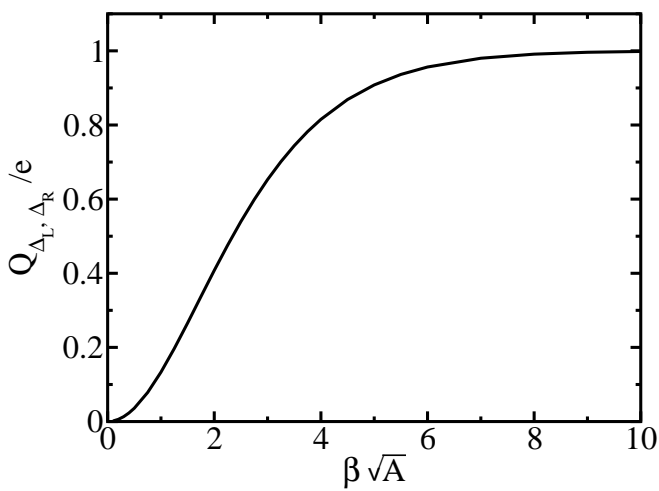}
\caption{Pumped charge per cycle as a function of the side of the square cycle in parameter space obtained by means of a numerical method. 
\label{fig8}}
\end{center}
\end{figure}

So far, we discussed the pumped charge in the weak-pumping limit. Now, we want to describe the transition from weak to strong pumping. Therefore, we increase the variation of the pumping parameters $\delta\Delta_j(t)$ and, hence, the area enclosed in the parameter space.
The pumped charge over a large pumping cycle can be obtained by integrating the weak-pumped charge of Eq.~(\ref{charge_exp}) over the area of the cycle. For the sake of definiteness, we choose as a pumping cycle  a square loop around the origin, as shown in Fig.~\ref{fig6}. The result for the charge as a function of the side of the loop,~$\sqrt{A}$, obtained in this way is plotted 
in Fig.~\ref{fig8}. 
In this figure we see that we obtain charge $e$ if the size of the cycle is larger than $k_\text{B} T$. This agrees with the results of the experiment of Pothier {\it et al.}\cite{pothier92} where charge quantization was measured for strong adiabatic pumping. 
Finally, we wish to stress that by means of our technique we are able to describe the transition from weak to strong pumping, recovering charge quantization for strong pumping cycles encircling the degeneracy point ${\bar{\Delta}}_\text{L}={\bar{\Delta}}_\text{R}=0$.

\section{Conclusions}

We developed a diagrammatic real-time approach to adiabatic pumping through a system of tunnel-coupled metallic islands, performing a systematic perturbative expansion in powers of
the tunnel-coupling strengths. This method allowed us to identify the
different physical processes which contribute to the pumped charge.
We first applied our formalism to a single-island system consisting
of one metallic island with Coulomb interaction tunnel coupled to two
noninteracting leads. We computed the pumped charge up to first order
in the tunnel-coupling strength, finding that the contribution in first
order is due to the renormalization of the charging-energy gap. For
the case of pumping with the two tunnel-coupling strengths, this term
becomes the dominant one. 
We emphasize that the single-island system works like a peristaltic pump if one tunnel-coupling strength and the charging-energy gap are the pumping parameters. For pumping with both tunnel-coupling strengths the peristalting mechanism occurs by means of charging-energy gap renormalization, therefore enabling experimental access  to renormalization effects. 

Furthermore, we considered pumping with
the charging-energy gaps in a system consisting of two tunnel-coupled
metallic islands. We calculated the pumped charge through this double-island system
and studied the issue of pumped-charge quantization.
First, we investigated the pumped charge in the weak-pumping limit. 
For example, we found that in experiments a symmetric choice of the average of the charging-energy gaps should be preferred to maximize the pumped charge.
By integrating the weak-pumped charge we obtained the pumped charge over a larger pumping cycle. Therefore, we were able to describe the transition from weak to strong pumping. In the strong-pumping limit we found the charge to be quantized. This is consistent with the results obtained by Pothier~{\it et~al.}\cite{pothier92} for strong pumping with the single-electron pump.

We acknowledge financial support from DFG via SPP 1285.

\begin{appendix}

\section{Examples of diagrams for the single-island system}
\label{appendix_A}

In this appendix, we give the results for the matrix elements $(\mathbf{W}_{t})_{i,j}$ for the single-island system. These matrices are needed to compute the occupation probabilities of the metallic island.
\begin{figure*}
\begin{center}
\includegraphics[width=1.\textwidth,clip]{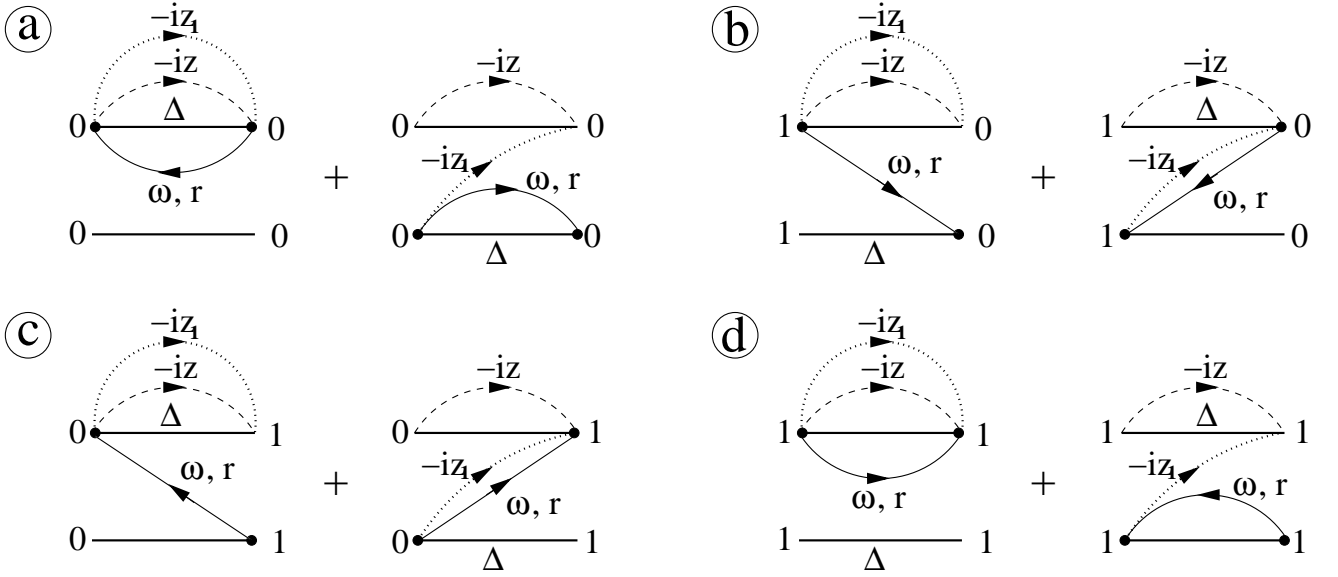}
\caption{(a) $(\mathbf{W}^{(\text{i/a},1)}_{t})_{0,0}\;$, 
(b) $(\mathbf{W}^{(\text{i/a},1)}_{t})_{0,1}\;$, 
(c) $(\mathbf{W}^{(\text{i/a},1)}_{t})_{1,0}\;$, and
(d) $(\mathbf{W}^{(\text{i/a},1)}_{t})_{1,1}\;$. 
Diagrams contributing to the instantaneous/ adiabatic matrix elements in first order in $\alpha_0$. The dotted frequency lines are only needed for the adiabatic corrections $(\mathbf{W}^{(\text{a},1)}_{t})_{i,j}$.
\label{fig9}}
\end{center}
\end{figure*}
We start with calculating the matrix elements $(\mathbf{W}_{t}^{(\text{i},1)})_{i,j}$ of the instantaneous kernels in first order in $\alpha_0$.
The corresponding diagrams are shown in Fig.~\ref{fig9}, where the additional dotted frequency lines must not be taken into account. 
As an example, we compute the two diagrams contributing to the matrix element $(\mathbf{W}_{t}^{(\text{i},1)})_{0,1}$, shown in Fig.~\ref{fig9}(b). 
The sum of the two diagrams reads 
\begin{eqnarray}
\label{W_i,1}
\left( \mathbf{W}_t^{(\text{i},1)}\right)_{0,1} 
&=& -2 \Im \left[ \sum_{r}\int  d\omega \dfrac{\alpha_r^-(\omega)}{\Delta-\omega+i0_+} \right] \nonumber \\
&=& 2 \pi \,\alpha^-(\Delta)\;.
\end{eqnarray}
The instantaneous kernel in first order in $\alpha_0$ reads
\renewcommand{\arraystretch}{2}
\begin{equation*}
 \mathbf{W}_t^{(\text{i},1)}
= 2 \pi 
\left( 
\begin{array}{cc}
-\alpha^+(\Delta)	&	\quad\alpha^-(\Delta)\\
\quad\alpha^+(\Delta)	&	- \alpha^-(\Delta)
\end{array}
\right) \;.
\end{equation*}
For the derivative of the instantaneous kernel in first order in $\alpha_0$ we get 
\renewcommand{\arraystretch}{2}
\begin{equation*}
\partial \mathbf{W}_t^{(\text{i},1)}
 = \alpha_0 \;\partial_\Delta\;\Re\left[ \varrho(\Delta)\right] 
\left( 
\begin{array}{cc}
\quad1 &	-1 \\
-1 	&	\quad1 
\end{array}
\right) \;.
\end{equation*}
To compute the adiabatic correction we need to introduce additional external frequency lines according to rules~(7a) and~(8a), which are drawn as dotted lines  in Fig.~\ref{fig9}. 
The evaluation of these diagrams leads us to the adiabatic correction to  kernel in first order in $\alpha_0$,
\renewcommand{\arraystretch}{2}
\begin{equation*}
\mathbf{W}_t^{(\text{a},1)}
 =\dfrac{d}{dt} \left\lbrace 
\dfrac{\alpha_0^{}}{2}\;  \partial_\Delta\;\Re\left[ \varrho(\Delta)\right] 
\right\rbrace 
\left( 
\begin{array}{cc}
\quad 1 &	-1 \\
-1 &	\quad 1 
\end{array}
\right) \;.
\end{equation*}
\begin{figure*}
\includegraphics[width=1.\textwidth,clip]{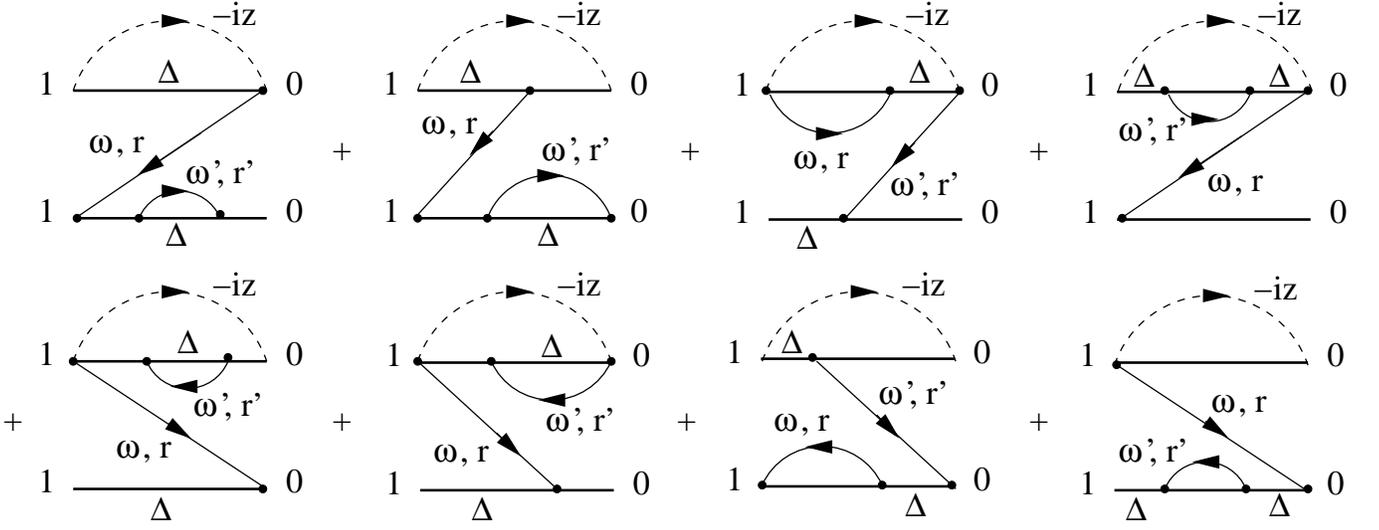}
\caption{Diagrams contributing to the instantaneous matrix element $(\mathbf{W}_{t}^{(\text{i},2)})_{0,1}$ in second order in $\alpha_0$. 
\label{fig10}}
\end{figure*}

Finally, we calculate the second-order contribution in $\alpha_0$ to the matrix elements of the instantaneous kernel. As an example, we show all diagrams contributing to the matrix element $(\mathbf{W}_{t}^{(\text{i},2)})_{0,1}$ in Fig.~\ref{fig10}. We have to sum over all indices $r, r'$ and to integrate over all frequencies $\omega, \omega'$. Performing the limit~$z=0_+$ and using  Cauchy's principal value, we get
\begin{eqnarray*}
\left( \mathbf{W}_t^{(\text{i},2)}\right)_{0,1} \;
&=& - \pi \, \alpha_0 \;  
\alpha(\Delta) \;   
\partial_\Delta\; \Re\left[ \varrho(\Delta)\right] \\ 
&\quad& + 2 \pi \,  \alpha_0 \; 
\Re\left[ \varrho(\Delta)\right] \;\;
\partial_\Delta \;\alpha^-(\Delta)   \,.
\end{eqnarray*}
For the matrix elements of the instantaneous kernel in second order in $\alpha_0$ we find
\renewcommand{\arraystretch}{2}
\begin{eqnarray*}
\mathbf{W}_t^{(\text{i},2)}
 &=&\quad \pi \, \alpha_0  \;\alpha(\Delta) \;  \partial_\Delta\;\Re\left[ \varrho(\Delta)\right] 
\left( 
\begin{array}{cc}
\quad1 &	-1 \\
-1 	&	\quad1 
\end{array}
\right) \\
&\quad&+ 2\pi \, \alpha_0 \; \Re\left[ \varrho(\Delta)\right]\;\partial_\Delta 
\left( 
\begin{array}{cc}
-\alpha^+(\Delta) &	\quad\alpha^-(\Delta) \\
\quad\alpha^+(\Delta) 	&	-\alpha^-(\Delta) 
\end{array}
\right) \;.
\end{eqnarray*}

\section{Examples of diagrams for the double-island system}
\label{appendix_B}

In this appendix we give the results for the matrix elements $(\mathbf{W}^{(\text{i},1)}_{t})_{i,j}$  of the instantaneous kernel in first order in $\alpha_0$ for the double-island system. 
\begin{figure*}
\begin{center}
\includegraphics[width=1.\textwidth,clip]{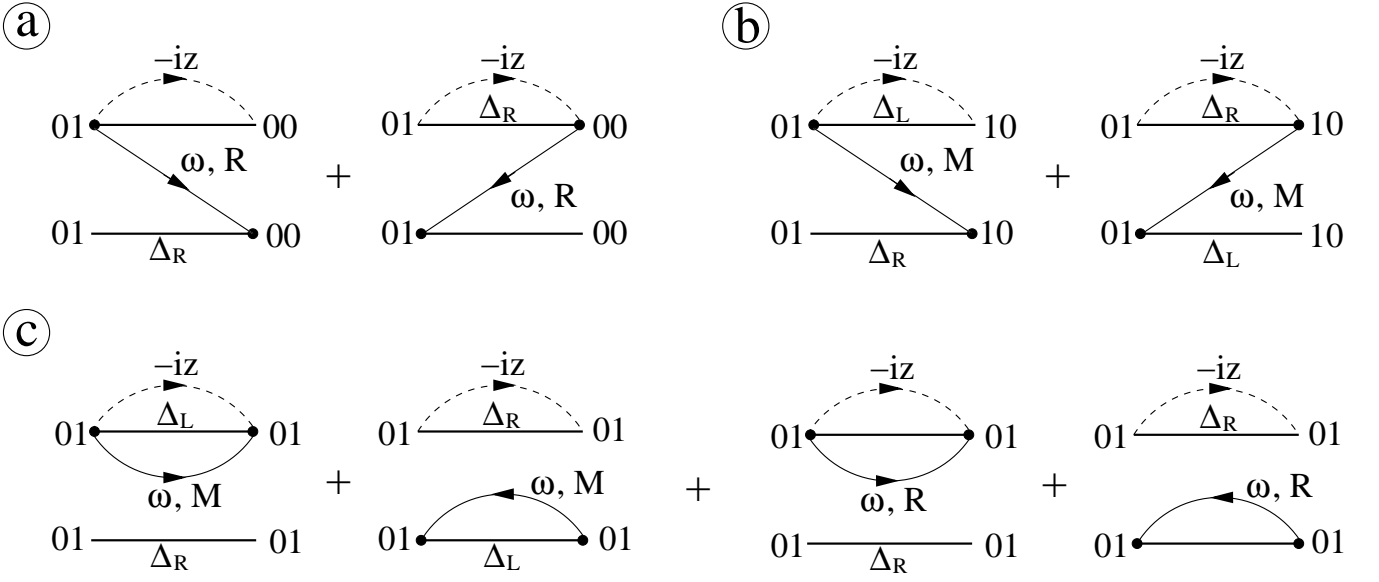}
\caption{(a) $(\mathbf{W}^{(\text{i},1)}_{t})_{00,01}\;$, 
 (b) $(\mathbf{W}^{(\text{i},1)}_{t})_{10,01}\;$,  and 
 (c) $(\mathbf{W}^{(\text{i},1)}_{t})_{01,01}\;$. 
Diagrams contributing to the instantaneous matrix elements in first order in $\alpha_0$. 
\label{fig11}}
\end{center}
\end{figure*}
Examples of the corresponding diagrams are shown in Fig.~\ref{fig11}. Their evaluation is similar to the one for the instantaneous matrix elements in first order in $\alpha_0$ for the single-island system, see Eq.~(\ref{W_i,1}),
\begin{eqnarray*}
\left( \mathbf{W}_t^{(\text{i},1)}\right)_{10,01} 
&=&-2 \Im \left[ \int {d\omega} 
\dfrac{\alpha_\text{M}^-(\omega)}{\Delta_\text{M}-\omega+i0_+}\right]\\ 
&=&2\pi\alpha_\text{M}^-(\Delta_\text{M}) \;.
\end{eqnarray*}

Proceeding in the same way for the other matrix elements and using the notation $\alpha_{r}^{\pm}=\alpha_{r}^{\pm}(\Delta_s)$ with $s=\left\lbrace \text{L},\text{R},\text{M} \right\rbrace$, we obtain for the matrix for the instantaneous kernel in first order in $\alpha_0$, 
\renewcommand{\arraystretch}{2}
\begin{equation*}
\mathbf{W}_t^{(\text{i},1)}
= 2 \pi 
\left( 
\begin{array}{ccc}
-\;\alpha_\text{L}^+ -\;\alpha_\text{R}^+	& \alpha_\text{L}^- 			& \alpha_\text{R}^- \\
\alpha_\text{L}^+			& -\;\alpha_\text{M}^+ -\;\alpha_\text{L}^- 	& \alpha_\text{M}^-\\
\alpha_\text{R}^+			& \alpha_\text{M}^+ 			& -\;\alpha_\text{M}^- -\;\alpha_\text{R}^-
\end{array}
\right) \;.
\end{equation*}

\end{appendix}

\end{document}